\documentstyle[11pt,newpasp,twoside,epsf]{article} 
\def\edcomment#1{\iffalse\marginpar{\raggedright\sl#1\/}\else\relax\fi} 
\reversemarginpar 

\begin{document} 
\title{UVES Abundances of Stars in Nearby Galaxies: 
How Good are Theoretical Isochrones?}

\author{Eline Tolstoy} 
\affil{Kapteyn Institute, University of Groningen, the Netherlands} 

\begin{abstract} 
Here we report some results from an ESO-VLT programme to observe
individual stars in nearby dwarf galaxies at high resolution with the
UVES spectrograph (Tolstoy, Venn, Shetrone, Primas, Hill, Kaufer \&
Szeifert 2002, submitted to AJ).  We mainly concentrate on
illustrating the issues and uncertainties surrounding our efforts
to determine the ages of stars for which we have accurately measured
[Fe/H] and [$\alpha$/Fe].  
\end{abstract}

\section{Determining Ages from Isochrones}
In principle it ought to be a straight forward task to plot the
theoretical isochrones at all ages (2$-$15~Gyr) for a star with an
accurately determined metallicity, and find the best fitting age.  We
have found that isochrones of different groups do not always produce
the same results, and on occasion none of the isochrones of any group
go through the position of the star in a Colour-Magnitude Diagram.
This is not found to be such a major problem for globular clusters.
To highlight the issues we plot in Figure~1 the results for three
stars in one of the four dwarf spheroidal galaxies in our sample -
Fornax, along with the Yale-Yonsei ($\alpha$=0) isochrones
of Yi {\it et al.} (2001, ApJS, 136, 417) at 2~Gyr \& 15~Gyr.
In one case the isochrones are too red, one too
blue, and one has no apparent problem.

Even though we reveal some worrying problems, in general we are
confident that we can attach ages to the stars we observe
which may lack some accuracy in absolute value, but they are appear to
be accurate relative to each other. We find that stars which
are more metal poor are older than the more metal rich stars. This
combination of metallicity and Colour-Magnitude Diagram analysis is
the only way to be able to disentangle the star formation history and
the corresponding chemical evolution in nearby galaxies.

This work highlights the difficulties in using theoretical stellar
evolution tracks verified to work for globular clusters to interpret
ages of stars in galaxies. This is not a particular problem of the
Yi et al. isochrones. This might be due to fundamental
differences in the formation and/or evolution of stars between a
globular cluster environment and that of field stars in a galaxy. It
is well known that stars in globular cluster stars 
contain different abundance patterns from all field stars.  We 
need a lot more high resolution spectra of stars in nearby galaxies to
better quantify the problem, because there is cuurently
no straight forward solution.

{\bf Acknowledgements:}
I thank the Royal Netherlands Academy of Arts and Sciences for a research
fellowship and travel support to attend this meeting.

\begin{figure}
\plotone{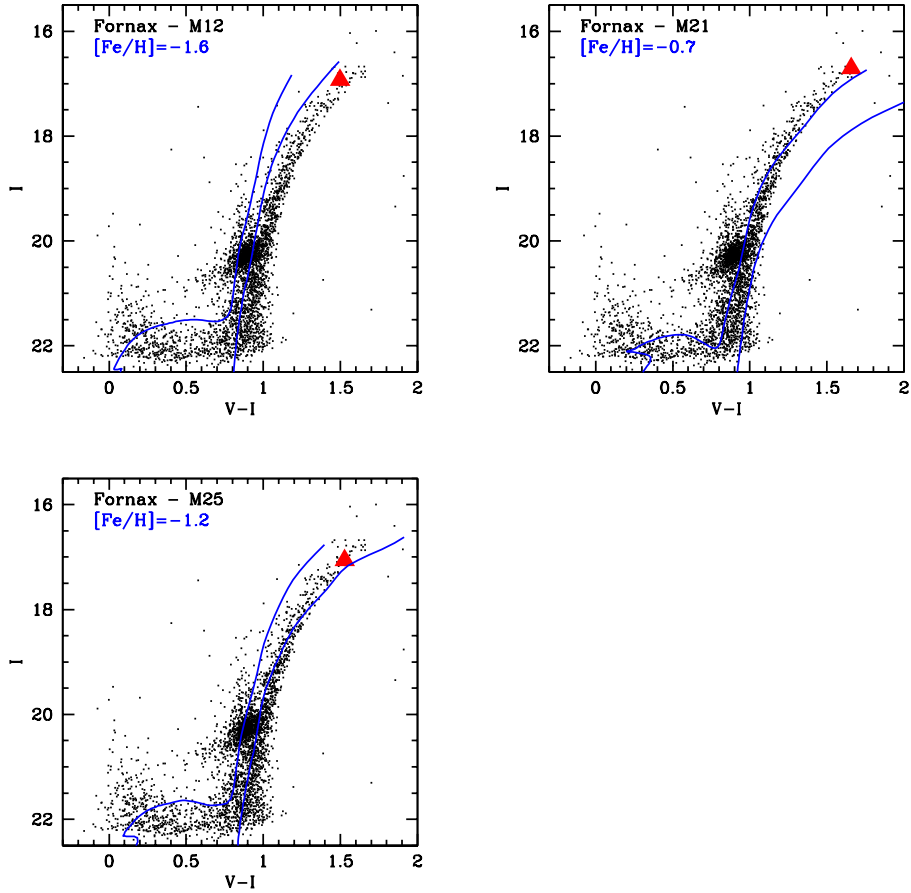}
\caption{Here we show the results of comparing UVES abundances measured for
stars in the Fornax Dwarf Spheroidal Galaxy.  In top left hand corner
of each plot is listed the name of the star and the UVES measured
[Fe/H] for the star symbol plotted on the Colour-Magnitude Diagram.
The Yale-Yonsei isochrones ($\alpha=0$) 
of Yi {\it et al.} (2001) at 2~Gyr \& 15~Gyr old are also plotted.
}
\end{figure}

\end{document}